\documentclass[showpacs,prl,twocolumn]{revtex4}
\usepackage{amsfonts}
\usepackage{amsmath}
\usepackage{amssymb}
\usepackage{graphicx}

\begin{document}

\title{Cavity-Mediated Strong Matter Wave Bistability in a Spin-1 Condensate}
\author{Lu Zhou$^{1}$, Han Pu$^{2}$, Hong Y. Ling$^{3}$, and Weiping Zhang$%
^{1}$}
\affiliation{$^{1}$State Key Laboratory of Precision Spectroscopy, Department of Physics,
East China Normal University, Shanghai 200062, China}
\affiliation{$^{2}$Department of Physics and Astronomy, and Rice Quantum Institute, Rice
University, Houston, TX 77251-1892, USA }
\affiliation{$^{3}$Department of Physics and Astronomy, Rowan University, Glassboro, New
Jersey 08028-1700, USA}

\begin{abstract}
We study matter wave bistability in a spin-1 Bose-Einstein condensate
dispersively coupled to a unidirectional ring cavity. A unique feature is
that the population exchange among different modes of matter fields are
accomplished via the spin-exchange collisions. We show that the interplay
between the atomic spin mixing and the cavity light field can lead to a
strong matter wave nonlinearity, making matter wave bistability in a cavity
at the single-photon level achievable.
\end{abstract}

\maketitle

The macroscopic nonlinear phenomena associated with ultracold atoms have
become a main stream of research interest in the emerging field of atom
optics \cite{book}. Meanwhile they establish intimate connection of the new
field to other branches of physics, such as nonlinear optics and condensed
matter physics. One important aspect in the field is how to coherently
manipulate the nonlinear behaviors of ultracold atomic ensembles for
different purposes.

The ability of an optical cavity to provide feedback between input and
output light fields can result in the modification of the atom-photon
interaction in a highly nonlinear fashion. The exploration of such a
nonlinear interaction for applications of both applied and fundamental
interest has led to many exciting developments, including optical
bistability, which was a subject of extensive study by the optics community
in the 1980's, due mostly to the prospect of its use as an optical switch in
all-optical computers \cite{gibbs85}. In recent years, rapid technological
advancement in cooling, trapping, and condensation of neutral atoms has
brought new opportunities to cavity quantum electrodynamics (QED). A
combination of cold atoms and large coherence couplings enables single-atom
trajectories to be monitored in real time with high signal-to-noise ratio
\cite{kimble98} while allows the vacuum Rabi splitting of a single trapped
atom to be experimentally observed \cite{kimble04}.

Instead of a \emph{single} atom, more recent studies in cavity QED focus on
cavity systems with a \emph{collection} of ultracold atoms \cite%
{esslinger07,esslinger08,kurn07,reichel07,kurn08,courteille07}, in which
strong coupling of ultracold atomic gases to cavity optical field are
realized. This allows us to enter a new regime of cavity QED, where a cavity
field at the level of a single photon can significantly affect the
collective motion of the atomic samples. This opens up new possibilities to
manipulate the nonlinear dynamics of ultracold atomic gases with
cavity-mediated nonlinear interaction.

So far the works involving in cavity with ultracold atomic gases have mainly
focused on the interplay between the cavity field and the atomic external
degrees of center-of-mass motion \cite%
{esslinger08,kurn07,lewenstein08,ritsch00,wumin,moore99,kurn08,courteille07}%
. The role of internal spin degrees of ultracold atomic gases in the
atom-cavity coupling has not yet been seriously explored. An intriguing
property of spinor Bose-Einstein condensate (BEC) is that in addition to the
repulsive binary collisions, atoms of different spin components can couple
to each other via spin-exchange interactions, which give rise to spin mixing
\cite{ho98}, a nonlinear dynamical phenomenon under intense theoretical \cite%
{pu98,you05,passos04} and experimental investigation \cite%
{ketterle981,hirano04,sengstock04,chapman05,lett07}. In this work we propose
a scheme to exploit the atom-cavity coupling to control the atomic spinor
dynamics in a spinor BEC. In contrast to the existing works which focus
primarily on optical bistability, here we pay particular attention to the
matter-wave bistability. As we will show, the combination of cavity-induced
phase shift and the intrinsic spin-exchange interaction of a spinor BEC
leads to very strong matter-wave bistability, providing a new playground for
exploring the spinor dynamics and cavity nonlinear optics.

\begin{figure}[tbh]
\includegraphics[width=8cm]{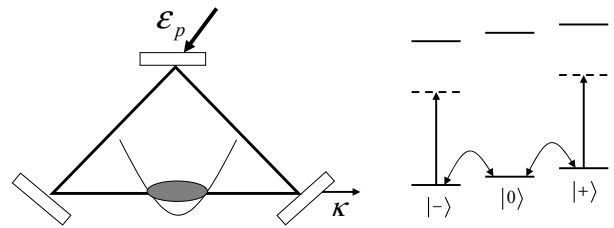}
\caption{{\protect\footnotesize Schematic diagram showing the system under
consideration.}}
\label{fig1}
\end{figure}

Our model --- a spinor BEC with hyperfine spin $F=1$ confined in a
unidirectional ring cavity --- is depicted schematically in Fig.~\ref{fig1}.
At zero temperature we assume single-mode approximation (SMA) that atoms in
different spin states can be described by the same spatial wave function $%
\phi \left( \mathbf{r}\right) $, then each spin component can be associated
with an annihilation operator $\hat{c}_{\alpha }$ ($\alpha =\pm $, $0$). A
weak external magnetic field may be applied to break the degeneracy and
provide the quantization axis. The cavity is designed in a way that only a
single traveling mode with frequency $\omega _{c}$, described by an
annihilation operator $\hat{a}$, interacts with the atoms. The cavity is
driven by a coherent laser field with frequency $\omega _{p}$ and amplitude $%
\varepsilon _{p}$.

The Hamiltonian under the SMA can be written as
\begin{equation*}
\hat{H}=\hat{H}_{0}+\left[ U_{0}\left( \hat{c}_{+}^{\dagger }\hat{c}_{+}+%
\hat{c}_{-}^{\dagger }\hat{c}_{-}\right) -\delta _{c}\right] \hat{a}%
^{\dagger }a+i\varepsilon _{p}\left( \hat{a}^{\dagger }-\hat{a}\right) ,
\end{equation*}%
with $\hat{H}_{0}$ describing the dynamics of spinor condensate
\begin{align*}
\hat{H}_{0}& =\lambda _{a}\left( \hat{c}_{+}^{\dagger }\hat{c}_{+}^{\dagger }%
\hat{c}_{+}\hat{c}_{+}+\hat{c}_{-}^{\dagger }\hat{c}_{-}^{\dagger }\hat{c}%
_{-}\hat{c}_{-}+2\hat{c}_{0}^{\dagger }\hat{c}_{0}\hat{c}_{+}^{\dagger }\hat{%
c}_{+}\right. \\
& \left. +2\hat{c}_{0}^{\dagger }\hat{c}_{0}\hat{c}_{-}^{\dagger }\hat{c}%
_{-}-2\hat{c}_{+}^{\dagger }\hat{c}_{+}^{\dagger }\hat{c}_{-}^{\dagger }\hat{%
c}_{-}+2\hat{c}_{0}^{\dagger }\hat{c}_{0}^{\dagger }\hat{c}_{+}\hat{c}_{-}+2%
\hat{c}_{+}^{\dagger }\hat{c}_{-}^{\dagger }\hat{c}_{0}\hat{c}_{0}\right) \\
& +q\left( \hat{c}_{+}^{\dagger }\hat{c}_{+}+\hat{c}_{-}^{\dagger }\hat{c}%
_{-}\right) ,
\end{align*}%
here $\lambda _{a}$ is the spin-dependent interaction coefficient \cite{ho98}
of the condensate. We denote $q$ as the quadratic Zeeman shift. $\delta
_{c}=\omega _{p}-\omega _{c}$ is the cavity-pump detuning. $%
U_{0}=g^{2}/\left( \omega _{p}-\omega _{a}\right) $, with $g$ being the
dipole coupling strength and $\omega _{a}$ the atomic transition frequency,
characterizes the strength of the atom-photon coupling. We will assume that
the photon frequency is sufficiently detuned away from the atomic
transitions so that the atomic upper level can be adiabatically eliminated
and the interaction between photon and atom is essentially of dispersive
nature. We assume that the photons are $\pi $ polarized which couple the
atoms in the $F=1$ ground-state manifold to the excited manifold with $%
F^{\prime }=1$. The transition selection rule is $\Delta m_{F}=0$. However,
since the transition $\left\vert F=1,m_{F}=0\right\rangle _{g}\rightarrow
\left\vert F^{\prime }=1,m_{F}^{\prime }=0\right\rangle _{e}$ is forbidden,
spin-0 level is not coupled by the photon. For simplicity, we also assume
that the coupling strength between the cavity field and spin-$\pm $ atoms
are the same. We will treat the leakage of cavity photons phenomenologically
by introducing a decay rate $\kappa $ with typical values $\sim 1$ MHz. By
contrast, the time scale for the atomic spin-mixing dynamics is much longer
--- the measured population oscillation frequency is below $10$ Hz for $%
^{87} $Rb \cite{chapman05} and around $50$ Hz for $^{23}$Na \cite{lett07}.
This separation of time scales allows us to assume that the cavity field
always follows adiabatically the atomic dynamics%
\begin{equation}
\hat{a}=\frac{\varepsilon _{p}}{\kappa -i\left[ \delta _{c}-U_{0}\left( \hat{%
c}_{+}^{\dagger }\hat{c}_{+}+\hat{c}_{-}^{\dagger }\hat{c}_{-}\right) \right]
}.  \label{quantum cavity}
\end{equation}%
The corresponding Heisenberg equations of motion for the atomic field
operators read%
\begin{equation}
i\dot{\hat{c}}_{\pm }=\left[ \hat{c}_{\pm },\hat{H}_{0}\right] +U_{0}\hat{a}%
^{\dagger }\hat{a}\hat{c}_{\pm },\text{ }i\dot{\hat{c}}_{0}=\left[ \hat{c}%
_{0},\hat{H}_{0}\right] .  \label{heisenberg}
\end{equation}%
Combining Eqs. (\ref{quantum cavity}) and (\ref{heisenberg}), in the bad
cavity limit one can find the effective Hamiltonian $\hat{H}_{eff}$ which
satisfies $i\dot{\hat{c}}_{\alpha }=\left[ \hat{c}_{\alpha },\hat{H}_{eff}%
\right] $%
\begin{equation}
\hat{H}_{eff}=H_{0}-\frac{\varepsilon _{p}^{2}}{\kappa }\tan ^{-1}\left[
\frac{\delta _{c}-U_{0}\left( \hat{c}_{+}^{\dagger }\hat{c}_{+}+\hat{c}%
_{-}^{\dagger }\hat{c}_{-}\right) }{\kappa }\right] .
\label{effective hamiltonian}
\end{equation}

In the following we adopt a mean-field treatment by replacing the operators $%
\hat{a}$ and $\hat{c}_{\alpha }$ with the corresponding $\mathcal{C}$
numbers $\alpha =\left\langle \hat{a}\right\rangle $ and $c_{\alpha }=\sqrt{%
N_{\alpha }}\exp \left( -i\theta _{\alpha }\right) $, where $N_{\alpha }$
and $\theta _{\alpha }$ represent the number and phase of the bosonic field
for the particles in the spin component $\alpha $, respectively. We take
advantage of the existence of two conserved quantities: the total atomic
number $N=N_{+}+N_{-}+N_{0}$ and magnetization $M=N_{+}-N_{-}$, and simplify
our problem into the one described by two variables: the normalized
population in the spin-$0$ component $x=N_{0}/N$ and the relative phase $%
\theta =2\theta _{0}-\theta _{+}-\theta _{-}$. The mean-field counterpart of
the quantum effective Hamiltonian (\ref{effective hamiltonian}) read%
\begin{align}
\frac{H}{N\kappa }& =\bar{q}\left( 1-x\right) +\bar{\lambda}_{a}x\left[ 1-x+%
\sqrt{\left( 1-x\right) ^{2}-m^{2}}\cos \theta \right]  \notag \\
& +U\left( x\right) ,  \label{mean field hamiltonian}
\end{align}%
with $U\left( x\right) \equiv \eta ^{2}\tan ^{-1}\left[ \bar{U}_{0}\left(
1-x\right) -\bar{\delta}_{c}\right] /N$ and $m=M/N$ is the atomic
polarization. We have defined other dimensionless parameters as
\begin{equation*}
\bar{\lambda}_{a}=\frac{N\lambda _{a}}{\kappa },\,\bar{q}=\frac{q}{\kappa }%
,\,\bar{U}_{0}=\frac{NU_{0}}{\kappa },\,\eta =\frac{\varepsilon _{p}}{\kappa
},\,\bar{\delta}_{c}=\frac{\delta _{c}}{\kappa }.
\end{equation*}%
The equations of motion for $x$ and $\theta $ read
\begin{subequations}
\label{classical equation}
\begin{align}
\frac{dx}{d\tau }& =2\bar{\lambda}_{a}x\sqrt{\left( 1-x\right) ^{2}-m^{2}}%
\sin \theta ,  \label{3a} \\
\frac{d\theta }{d\tau }& =-2\left( \bar{q}+\frac{\bar{U}_{0}\left\vert
\alpha \right\vert ^{2}}{N}\right)  \notag \\
& +2\bar{\lambda}_{a}\left[ 1-2x+\frac{\left( 1-x\right) \left( 1-2x\right)
-m^{2}}{\sqrt{\left( 1-x\right) ^{2}-m^{2}}}\cos \theta \right] ,  \label{3b}
\end{align}%
where $\tau =\kappa t$ is the dimensionless time.

From Eq. (\ref{3b}) one can see that the cavity modifies the atomic dynamics
in the same manner as the quadratic Zeeman terms $\bar{q}$, which will lead
to redistribution of atomic population among different spin states through
spin mixing. However, this cavity-induced effective Zeeman energy is
dependent upon the atomic population distribution via Eq.~(\ref{quantum
cavity}). It is this inter-dependence of the atomic and photonic modes that
leads to interesting nonlinear dynamics of this coupled system, which will
be the focus of this work.

\begin{figure}[tbh]
\includegraphics[width=7cm]{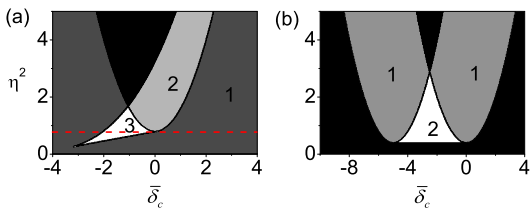}
\caption{{\protect\footnotesize (Color online) Phase diagram in the
parameter space of }$\bar{\protect\delta}_{c}${\protect\footnotesize \ and }$%
\protect\eta ^{2}${\protect\footnotesize \ for different type of solutions:
(a) }$\protect\theta =0${\protect\footnotesize ; (b) }$\protect\theta =%
\protect\pi ${\protect\footnotesize . Different regions are differentiated
by their colors and are labeled with the numbers of corresponding solutions.
In the black region, no physical phase-dependent solutions can be found. The
dimentionless parameters are estimated to be }$\bar{\protect\lambda}%
_{a}=10^{-3}${\protect\footnotesize , }$\bar{q}=2\bar{\protect\lambda}_{a}$%
{\protect\footnotesize \ and }$\bar{U}_{0}=-5${\protect\footnotesize \
\protect\cite{parameter}, the other parameters are set as }$m=0$%
{\protect\footnotesize \ and }$N=1000${\protect\footnotesize . The red
dashed line in (a) correspond to }$\protect\eta ^{2}=0.8$%
{\protect\footnotesize .}}
\label{fig2}
\end{figure}

The dynamics of the system can be captured by the contour plot of the
Hamiltonian $H$, which is intimately related to the fixed points $\left(
x_{0},\theta _{0}\right) $ given by the equilibrium solution of Eqs. (\ref%
{classical equation}). In this work, we only consider the anti-ferromagnetic
atoms ($^{23}$Na) with $\bar{\lambda}_{a}>0$. The ferromagnetic case is not
qualitatively different. In the absence of the cavity field, the equilibrium
solutions $\left( x_{0},\theta _{0}\right) $ have been studied in \cite%
{passos04,you03}. Besides the phase-independent solutions of $x_{0}=0$ and $%
x_{0}=1-|m|$ for which the relative phase $\theta _{0}$ is not well-defined,
the spinor condensate system supports at most one phase-dependent solution
with $\theta _{0}=0$ or $\pi $. The presence of the cavity field
dramatically changes this property. The phase diagram identifying different
types of solution is mapped out in the parameter space of $\eta ^{2}$ and $%
\bar{\delta}_{c}$, as shown in Fig.~\ref{fig2}. We can see that, in certain
parameter regime, the number of different phase-dependent solutions can be
more than one, different solution regimes of the coupling system can be
crossed by varying $\bar{\delta}_{c}$ and $\eta ^{2}$. Since these two
parameters are directly related to the pump laser, this means that the
dynamical properties of the system can be easily manipulated by tuning the
intensity or frequency of the pump laser field.

\begin{figure}[tbh]
\includegraphics[width=8cm]{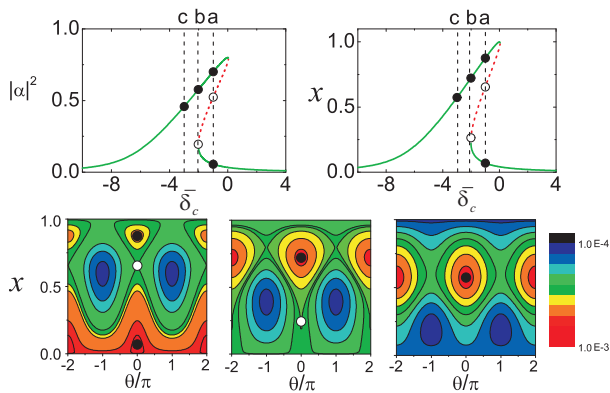}
\caption{{\protect\footnotesize (Color online) Upper panel: Mean intracavity
photon number }$\left\vert \protect\alpha \right\vert ^{2}$%
{\protect\footnotesize \ and the normalized spin-0 population }$x$%
{\protect\footnotesize \ versus cavity-pump detuning }$\bar{\protect\delta}%
_{c}${\protect\footnotesize \ for the steady-state solutions with $\protect%
\eta ^{2}=0.8$, corresponding to the red-dashed line in Fig.~\protect\ref%
{fig2}(a). The ones that represented by the red dotted lines correspond to
dynamically unstable solutions. Lower panel: From left to right, the
phase-space contour plot of }$H${\protect\footnotesize \ corresponding to
different values of }$\bar{\protect\delta}_{c}${\protect\footnotesize \
marked in the upper panel as a, b and c, respectively.}}
\label{bistability}
\end{figure}

Here we consider the case with the pump intensity $\eta ^{2}$ fixed, by
varying the cavity-pump detuning $\bar{\delta}_{c}$, the equilibrium
properties of the system are changed, as shown in the red-dashed line in
Fig. \ref{fig2}(a). The corresponding phase-dependent fixed points are
derived and the results are shown in Fig.~\ref{bistability}. The system
exhibits typical bistable behavior: For certain values of $\bar{\delta}_{c}$%
, it supports three stationary solutions. A standard linear stability
analysis shows that in the region with three solutions, two of these are
dynamically stable and the third one is dynamically unstable. Further
insights can be gained by examing the corresponding contour plot of $H$
(Fig.~\ref{bistability}, lower panel). The unstable fixed points correspond
to the saddle points in the contour plots.

The same hysteresis feature can also be identified from the mean-field
energy\ diagram by the appearance of a swallowtail loop structure (the dots
in Fig. \ref{energyspectrum}). \ A quantum calculation involving a direct
diagonalization of the effective Hamiltonian $\hat{H}_{eff}$\ confirms a
well-known correspondence between the semiclassical and quantum energy
levels \cite{quantum-classical}, namely, in the bistable region the quantum
spectrum (the solid lines in Fig. \ref{energyspectrum}) exhibits a series of
anticrossings and when connected, these anticrossings form the top segment
of the swallowtail structure of the corresponding mean-field energy level.
(In this example, we have, without loss of the essential physics, adopted a
much smaller system so that the quantum calculation can be done within a
reasonable computational time.)

\begin{figure}[tbh]
\includegraphics[width=5cm]{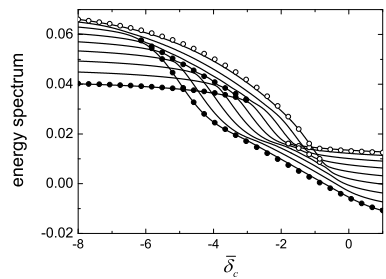}
\caption{{\protect\footnotesize Quantum and mean-field energy levels with }$%
N=20${\protect\footnotesize , }$m=0.2${\protect\footnotesize \ and }$\protect%
\eta ^{2}=0.02${\protect\footnotesize , the other parameters are the same as
before. The solid lines are quantum energy levels, the black dots refer to
the mean-field energy levels with }$\protect\theta =\protect\pi $%
{\protect\footnotesize , while the white dots refer to those with }$\protect%
\theta =0${\protect\footnotesize .}}
\label{energyspectrum}
\end{figure}

Let us now return to Fig. \ref{bistability}. As indicated in the upper panel
of Fig.~\ref{bistability}, both the cavity field and the atoms exhibit
bistable behavior. The mean cavity photon number involved is always less
than unity. Remarkably, such a small number of photons affect the whole
condensate and lead to complete population redistribution among different
internal atomic spin states, which can be readily observed in experiment.
This behavior can be understood as following: The collective nature of the
condensate greatly enhances the atom-photon coupling such that a single
photon gives rise to a significant atomic phase shift, which in turn
strongly modifies the population distribution among the spin states.
Bistability results from the nonlinear feedback between photons and atoms.

It is interesting to compare our study with the experimental work of Refs.~%
\cite{esslinger08,kurn07}, where the motional degrees of freedom of
ultracold atomic gases represent the source of nolinearity affecting
light-atom interactions. In their system the atomic zero-momentum mode and
two side modes with momentum $\pm 2\hbar k$\ ($k$\ is the cavity light wave
vector) dominate the dynamics. The nonlinearity originate from the coherence
between these modes which is induced by the coupling provided by the
standing-wave cavity field \cite{ling01}. Such a coupling is neither strong
as it is provided by the weak cavity field, nor resonant as there is a
detuning of $4\hbar \omega _{rec}$\ with $\omega _{rec}=\hbar k^{2}/2m$\ the
atomic recoil frequency.\textbf{\ }In these systems, optical bistability of
low photon numbers is possible and is indeed observed. In principle, atomic
population in different motional states should also exhibit bistability.
However, such matter-wave bistability will be difficult to observe as it is
not easy to measure atomic population of different momentum states inside a
cavity in real time. Furthermore, due to the inefficient coupling between
atomic momentum modes as we just mentioned, the matter-wave bistability is
very weak since most of the atomic population will remain in the
zero-momentum state. In a recent work \cite{wumin} where this system is
theoretically examined, it is found that bistable behavior involving tens of
photons can only transfer about 20\% of the total atomic population out of
the zero-momentum state (see Fig.~1 of Ref.~\cite{wumin}).

In contrast, in the model we considered here, the coherence between the
internal atomic spin states affecting atom-light interaction is induced by
the intrinsic spin-exchange interaction which represents a matter-wave
analog of the four-wave mixing in nonlinear optics. An immediate advantage
is that it can be independently tuned with respect to the cavity field,
thereby dramatically increasing the chance of large population change among
spin states at a low cavity field. Consequently, our system can exhibit very
strong matter-wave bistability. This is indeed confirmed by our detailed
calculations.

In summary, we have studied the mutual interaction of a spinor condensate
with a single-mode cavity field. We show that the coupled cavity-spinor
condensate system can display simultaneously strong optical bistability at
the single-photon level and strong matter-wave bistability involving a whole
condensate with macroscopic number of atoms. This opens up new opportunities
to explore a diversity of new phenomena in cavity nonlinear optics with low
photon numbers and many-body physics with quantum gases. Before ending, we
note that the condensate depletion may become significant in the long time
scale as the quantum fluctuations of the cavity can introduce excess noise
to the condensate system \cite{domokos}. A more careful treatment taking
these effects into proper account will be left for further investigation.

This work is supported by the National Natural Science Foundation of China
under Grant No. 10588402, the National Basic Research Program of China (973
Program) under Grant No. 2006CB921104, the Program of Shanghai Subject Chief
Scientist under Grant No. 08XD14017, Shanghai Leading Academic Discipline
Project under Grant No. B480 (W.Z.), and by the NSF (H.P., H.Y.L.), ARO
(H.Y.L.), and the Welch Foundation with grant C-1669 (H.P.).

\end{subequations}


\begin{thebibliography}{99}
\bibitem{parameter} The dimensionless parameters are estimated with
experimentally accessible parameters: $\kappa \sim 2\pi \times 20$ KHz \cite%
{courteille07} and $N\lambda _{a}\sim 2\pi \times 20$ Hz for sodium
atoms with a typical density $10^{14}$ cm$^{-3}$ \cite{lett07},
$q\sim 2\pi \times 40$ Hz and $U_{0}\sim -2\pi \times 100$ Hz.
\end{thebibliography}
\end{document}